\documentclass[12pt, epsfig]{article}

\usepackage{amsmath}

\usepackage[hmargin=2.4 cm,vmargin=2cm]{geometry}
\usepackage{graphicx}
\begin{document}
\title{Entropy of singularities in self-gravitating radiation}
\author {Charis Anastopoulos\footnote{anastop@physics.upatras.gr} \\
 {\small Department of Physics, University of Patras, 26500 Greece} \\
 and  \\ Ntina Savvidou\footnote{ntina@imperial.ac.uk} \\
  {\small  Theoretical Physics Group, Imperial College, SW7 2BZ,
London, UK} }
\maketitle

\begin{abstract}
The Bekenstein-Hawking entropy suggests that thermodynamics is an intrinsic ingredient of gravity. Here, we explore the idea that requirements of thermodynamic consistency could determine the gravitational entropy in other set-ups.   We implement this idea in a simple model: static, spherically symmetric solutions to Einstein's equations corresponding to self-gravitating radiation. We find that the principle of maximum entropy provides a consistent thermodynamic description of the system, only if the entropy includes a contribution from the spacetime singularities that appear in the solutions of Einstein's equations. The form of the singularity entropy is stringently constrained from   consistency requirements, so that the existence of a simple expression satisfying these constraints is highly non-trivial, and suggests of a fundamental origin.  We find that the system is characterized by three equilibrium phases, and we conduct a preliminary investigation of the associated phase transitions. These results   demonstrate the point that  gravitational entities  other than horizons are endowed with thermodynamic properties.
\end{abstract}

\section{Introduction}
The current classical theory of gravity is very successful both in providing a broad theoretical framework and in its agreement with experiments. However, in its present form it is also incomplete in the sense that (i) it does not include in one integrated picture the causal notion of singularities; and (ii) it does not contain an intrinsic notion of entropy that it is of crucial importance for the study of black hole thermodynamics. The latter issue has motivated several ideas about the integration of gravity and thermodynamics. Some interesting examples are Jacobson's interpretation of Einstein's equations as equations of state \cite{Jac}, Padmanabhan's programme on gravity as a thermodynamic theory \cite{Pad} and Verlinde's interpretation of gravity as an entropic force \cite{Ver}---see also Ref. \cite{Hu} for a related review.

At present, most   discussions on gravitational entropy focus on the entropy of horizons (black hole or particle horizons), for which precise mathematical expressions exist. However, horizons are not the only gravitational entities to which entropy can be assigned. In particular, Penrose has proposed the association of entropy with gravitational singularities as a way to obtain a thermodynamically consistent description of cosmology \cite{Pen}. A concrete expression for the entropy of singularities, which is currently missing (at least for singularities in Lorentzian space-times),
 would greatly increase the scope of the search for a general theory of gravitational entropy.

The aim of this work is to present  a concrete expression for the entropy associated to a class of spacetime singularities, using elementary  principles of general relativity and thermodynamics. This result provides a direct verification of the conjecture that (at least some of) the singularities   can be viewed as thermodynamic objects.

The key idea is the following. If the gravitational degrees of freedom are carriers of entropy, then the total entropy in a gravitating system should be the sum of matter entropy and gravitational entropy (plus a small term due to their interaction.) However, the form of any entropy functional is constrained by requirements of thermodynamic consistency. In particular, a well-defined entropy functional  should allow for the implementation of the {\em maximum-entropy principle}: in an equilibrium state, all unconstrained parameters describing a physical system take values that maximize the entropy functional. If one ignores the contribution of gravitational entropy,   the resulting entropy functional may fail to satisfy defining properties of entropy and thus, misidentify the physical equilibrium states. Hence, the requirement that the entropy functional  satisfies  basic thermodynamic consistency conditions could allow us to infer the form and properties of any contribution to entropy from the gravitational field.  We show that this idea is successfully implemented in static, spherically symmetric solutions to Einstein equations describing self-gravitating radiation \cite{PWZ, Cha1}. It leads to an expression for gravitational entropy that is defined in terms of the properties the spacetime singularities that generically appear in these solutions.

The key physical requirement in this paper is the  consistency of the thermodynamic description. It is, however, to be noted that the presence of a long-range force such as gravity leads to a significant modification of the usual thermodynamical properties---for reviews, see, Refs. \cite{Padm, Katz, Cha2, CDR}. Most important is the loss of  extensivity in the thermodynamic potentials. In  ordinary equilibrium  thermodynamics, the entropy is an extensive function on the system's thermodynamic state space $\Gamma$: $S(\lambda X) = \lambda S(X)$, for any $\lambda >0$ and $X \in \Gamma$. This is not the case in gravity, even though in certain systems other scaling properties exist \cite{Opp, Pes}. The loss of extensivity implies that: the equilibrium configurations are inhomogeneous; the heat capacity may become negative, even though the specific heat of each volume element is positive \cite{LL, Thi}; the microcanonical and the canonical distribution are not equivalent \cite{ineq, CDR}; and, crucially, matter entropy may not have global maxima for fixed values of energy \cite{gth}.

Thus,  many concepts of ordinary equilibrium thermodynamic concepts must be  modified in order to fit the gravitational context. Nonetheless,  black hole entropy suggests that the thermodynamic description constitutes a fundamental element of gravitational physics. For this reason,  the fundamental principles of thermodynamics, as encapsulated in its basic laws, can be expected to hold even in a less stringent form. In this regard, we note that many concepts of the axiomatic approaches to equilibrium thermodynamics---for example \cite{Gil,LY}---are applicable in the gravitational context. In these approaches, entropy is viewed as a function on a state space of macroscopic observables that determines the allowed transitions of a physical system, and thus it can be defined irrespective of whether the extensivity property is satisfied or not. In particular, the maximum-entropy principle, which is the key thermodynamic principle employed in this paper, is based on the statistical interpretation of entropy and on the second law of thermodynamics. These  are universal principles and they are expected to apply also to gravitating systems.

With these considerations in mind we revisit the thermodynamics
of static, spherically symmetric solutions to Einstein equations corresponding to radiation in a box.
 The stationarity assumption allows us to employ notions of equilibrium thermodynamics---entropy in the full gravity theory would require notions of non-equilibrium thermodynamics. Spherical symmetry implies that the local degrees of freedom of the gravitational field are frozen, so that any contribution to the gravitational entropy would be `topological'. In the present context, this means that gravitational entropy can be identified with the entropy of singularities---or possibly with entropy associated to internal boundaries. We chose the case of  self-gravitating radiation \cite{PWZ}, because its thermodynamic description is simple: the equation of state is linear and it contains no scale parameters.
This allows for a semi-analytic treatment of the system and a relatively straightforward determination of the gravitational contributions to entropy. Furthermore, the equation of state for radiation is expected to be physically meaningful even when the energy density varies over many orders of magnitude.

The basic thermodynamic variables for self-gravitating radiation in a box are the ones accessible to an observer outside the box, namely, the area $4 \pi R^2$ of the box, the local temperature $T$ of the box and the Arnowitt-Deser-Misner (ADM) mass $M$. The variables $R$, $M$ and $T$ define the boundary conditions of Einstein' equations for this system. They can be varied independently, each choice of  $(R, M, T)$ defining a different solution to Einstein's equations. However,  $R$, $M$ and $T$ are not independent, when viewed as thermodynamic variables. For fixed size of the bounding box, the temperature of radiation is a function of the total energy. Thus, states of thermal equilibrium satisfy  a {\em structure relation} of the form $f(R, M, T) = 0$, which cannot be determined by Einstein's equations alone.
The standard treatment of static spherically symmetric solutions to Einstein's equations introduces an additional assumption, of regularity at the center (see, Ref. \cite{PWZ, Cha1} for self-gravitating radiation). Regular solutions form a set of measure zero in the space of all solutions; most solutions have a conical singularity at the center. The restriction to regular solutions follows mainly from the desire to avoid gravitational singularities, but it also provides the additional condition that is necessary  in order to derive  the thermodynamic structure equation $f(R, M, T) = 0$.

However, the regularity condition has no thermodynamic justification. For example, it is not obtained from the extremization of any thermodynamic potential. If we assume that general relativity arises as the macroscopic limit of an underlying microscopic theory, the lack of a fundamental justification for the regularity condition poses a severe problem. There should be some physical mechanism guaranteeing the stability of regular solutions against statistical or quantum fluctuations of the microscopic degrees of freedom. At the macroscopic level, such a mechanism should be described in the language of thermodynamics.

Regular solutions exist only in a specific sector of the thermodynamic state space. Outside this sector, the regularity condition cannot identify the equilibrium states.  Again, this is   problematic from the thermodynamic point of view. To see this, we note that for regular solutions  there is an upper bound $T_{max}(R)$ on the temperature of the bounding box for a given value of $R$ \cite{PWZ, Cha1}. Suppose that we bring the box into contact with a   thermal reservoir at temperature $ T > T_{max}(R)$. One expects that the box also acquires temperature equal to $T$ \cite{oth}. But there exists no regular solution with this property, so we have no way to determine the resulting equilibrium configuration.    Thus, the regularity condition does not suffice to describe even elementary thermodynamic operations to the system.

In ordinary equilibrium thermodynamics, when the external constraints (boundary conditions) to a system do not suffice to determine the system's state, the maximum-entropy principle is invoked \cite{Callen}. However,  the entropy $S_{rad}$ of self-gravitating radiation does not have global maxima for fixed values of energy. Thus, the maximum-entropy principle does not recover the regular solutions, even at the limit of weak gravity. This is a severe consistency problem, because at this limit we expect the ordinary thermodynamics of a photon gas to provide an excellent approximation.

The problems above are resolved by the assumption that an entropy $S_{sing}$ is associated to the
 singularities characterizing the solutions to Einstein's equations. $S_{sing}$ is defined only by the asymptotic behavior of the fields near the singularity, and its form is determined by the requirement
 that the total entropy $S_{tot} = S_{rad} + S_{sing}$ is maximized at the regular solutions.  Thus, {\em the regularity condition arises as a consequence of the maximum-entropy principle}. The maximum-entropy principle also determines the equilibrium configurations in   the sector where no regular solutions exist.   We find two phases of singular equilibrium solutions in addition to the phase of regular solutions.

It is important to emphasize that there are   strong constraints to the form of $S_{sing}$, arising from requirements of thermodynamic and mathematical consistency. These constraints are so stringent, that it is {\em a priori} unlikely that they can be satisfied accidentally, or even by fine-tuning. Hence, the existence of an expression for $S_{sing}$ that satisfies the constraints is highly remarkable. It is even more so, because  $S_{sing}$ has a very  simple form, and there are plausible  arguments suggesting that it  is unique. For this reason, we believe that the expression for  $S_{sing}$, determined here, is of a fundamental origin and it extracts information that is deeply embedded into the structure of Einstein's equations. Further elaboration of these ideas into other systems could provide significant novel information towards a general theory of gravitational entropy.

The association of an entropy $S_{sing}$ to the singularities does not necessarily imply a commitment into the physical existence of singularities. As we explain in more detail in Sec. 3.2, we view $S_{sing}$ as a mathematical term, defined by the asymptotic field values on the singularity, that is included in the total entropy  in order to obtain a consistent thermodynamic description of the system.  The expressions "entropy of singularity" and "singularity entropy" employed in this paper are to be understood in this sense. They do not imply the existence of the singularity as a physical entity that possesses entropy.

The definition and properties of the entropy $S_{sing}$  is consistent with the idea of an intrinsic thermodynamic character of gravity. Borrowing Padmanabhan's analogy of spacetime to an elastic solid \cite{Pad}, the singularities can be viewed as analogous to defects in the solid. For weak external stress, the solid's equilibrium state is smooth and regular, but if the stress increases beyond a critical value, no globally regular configuration exists:  configurations with defects are entropically favored. The thermodynamic picture of gravity provides an important shift of perspective: the singularities are not pathologies in the description of the system, but they may  indicate a different thermodynamic `phase' that   is entropically favored for certain boundary conditions.

\smallskip

The structure of this paper is the following. In Sec. 2, we study in detail the properties of the singular solutions of the Oppenheimer-Volkoff equations for self-gravitating radiation in a spherical box. In Sec. 3, we define the singularity entropy $S_{sing}$ and we show that the regular solutions are indeed obtained by a maximum-entropy principle. We also identify the phase of singular equilibrium solutions and study the corresponding phase transitions. In Sec. 4 we discuss our results and their implications.

\section{Solutions of the Oppenheimer-Volkoff equations for self-gravitating radiation}

\subsection{Structure equations and boundary conditions}

The thermodynamic system under consideration is a spherical box of area $A = \pi R^2$ containing a self-gravitating photon gas in thermal equilibrium. The local temperature $T$ of the photon gas is determined by the energy density $\rho$,
\begin{eqnarray}
\rho = \sigma T^4, \label{SB}
\end{eqnarray}
where $\sigma$ is the
 Stefan-Boltzmann constant.  The equation of state
 is $P = \frac{1}{3} \rho$, where $P$ is the pressure.

  In what follows, we work entirely within the confines of the classical theory, so $\hbar$ does not appear. It is therefore convenient to choose units where  $\sigma = 1$.

Assuming that the system is in equilibrium, we consider only static solutions to Einstein's equations. We  denote the associated timelike Killing field as $\frac{\partial}{\partial t}$.
 Outside the box, the spacetime metric is the Schwarzschild solution with  ADM mass $M$. Inside the box, the metric is of the form
\begin{eqnarray}
ds^2 = - (1 - \frac{2M}{R}) \sqrt{\frac{\rho(R)}{\rho(r)}} dt^2 + \frac{dr^2}{1 - \frac{2m(r)}{r}} + r^2 (d\theta^2 + \sin^2\theta d \phi^2), \label{ds2}
\end{eqnarray}
in the usual spherical-symmetric coordinates $(t, r, \theta, \phi)$. The mass function $m(r)$ satisfies the equation
\begin{eqnarray}
\frac{dm}{dr} = 4 \pi r^2 \rho, \label{dm}
\end{eqnarray}
with the boundary condition $m(R) = M$. The energy density $\rho(r)$ is obtained by the solution of the Oppenheimer-Volkoff equation
\begin{eqnarray}
\frac{d \rho}{dr} = -\frac{4 \rho}{r^2} \frac{(m+ \frac{4}{3} \pi r^3 \rho)}{1 - \frac{2m}{r}}. \label{OV}
\end{eqnarray}

An external observer has access only to variables that can be measured outside the box, namely, the box's area $A$, its local temperature $T$ and the ADM mass $M$. (The mass is determined by measurements of the  acceleration of free-falling test particles near the box).  By Eq. (\ref{SB}), the temperature $T$   determines the value of the density at the boundary $\rho(R)$. Thus, the knowledge of $M = m(R)$ and $T = [\rho(R)]^{1/4}$ allows us to integrate the system of Eqs. (\ref{dm}) and (\ref{OV}) from the boundary $r = R$ towards the center $r = 0$, and thus identify the spacetime geometry inside the box.

The system of Eqs. (\ref{dm}--\ref{OV}), integrated from the boundary $r = R$ inwards, admits two different classes of solutions,  distinguished by their behavior as $r \rightarrow 0$. The generic case corresponds to  solutions with a conical singularity at $r = 0$. In these solutions,  $m(0) = -M_0$ for some positive constant $M_0$, and $\rho \sim r^2$ as $r \rightarrow 0$. The other class consists of regular solutions; these satisfy $m(0) = 0 $ and $\rho(0) = \rho_c $ for some constant $\rho_c >0$. The regular solutions form a set of measure zero in the space of all solutions. They define  spacetimes that are everywhere locally Minkowskian. In Sec. 2.3, we show that no solutions  with $m(0)   > 0$ exist when  Eqs. (\ref{dm}) and (\ref{OV}) are integrated from the boundary inwards.

In regular solutions, $m(r)$ is positive valued and $m(r \rightarrow \infty) = M$, hence, it is usually interpreted as the total energy within a spherical surface of radius $r$. This interpretation is heuristic, because in general relativity there is no invariant way of defining the total energy in a finite region. In singular, solutions, the function $m(r)$ takes negative values near the singularity so its interpretation as total energy is untenable. In particular, the value $m(0) = -M_0$ is a topological feature of the singularity, with no {\em a priori} relation to energy. As we explain in Sec. 3.3, a better measure of energy in singular solutions is obtained in terms of the Komar integral.

Energy and temperature are not independent variables. In an ordinary (i.e., non-gravitating) photon gas the entropy $S$ is a function of the energy $U$ and the volume $V$, and the temperature $T$ is defined as $T^{-1} = (\partial S/\partial U)_V$. Similarly, in self-gravitating radiation, there are only {\em two} independent thermodynamic variables (for example, $M$ and $R$). However, a unique solution of Eqs.  (\ref{dm}) and (\ref{OV}) requires the specification of {\em three} independent variables, $M, T$ and $R$.

In usual studies of spherical symmetric solutions to Einstein's equations, one restricts attention to the subspace of {\em regular} solutions. For self-gravitating radiation, this implies solving Eqs. (\ref{dm}) and (\ref{OV}) as a boundary-value problem, with $m(0) = 0$ and $m(R) = M$. Regularity is an additional condition that determines the functional dependence of the temperature $T$ on the variables $R$ and $M$.

The equation of state for radiation does not contain any length scale. For this reason,  the Oppenheimer-Volkoff equation possesses a scaling symmetry: the transformation $M\rightarrow \lambda m, R \rightarrow \lambda R, T \rightarrow \lambda^{-1/2}T$ preserves the solution curves of the system. This symmetry implies that the structure equations are equivalent to a two-dimensional time-homogeneous dynamical system. The definition of the variables
\begin{eqnarray}
\xi &=& \ln \frac{r}{R} \\
u &=& \frac{2m(r)}{r} \\
v &=& 4 \pi r^2 \rho
\end{eqnarray}
brings Eqs. (\ref{dm}) and (\ref{OV}) into the form
\begin{eqnarray}
u' &=& 2v - u \label{equ}\\
v' &=& \frac{2v (1 - 2 u - \frac{2}{3}v)}{1-u}, \label{eqv}
\end{eqnarray}
where the prime refers to differentiation with respect to $\xi$.  Eqs. (\ref{equ}) and (\ref{eqv}) are integrated from $\xi = 0 $ (boundary, $r = R$) to $\xi \rightarrow - \infty$ (center, $r = 0$). In what follows, we denote the values of $u$ and $v$ at the boundary as $u_0 := u(\xi = 0)$ and $v_0 := v(\xi =0)$. Thus, $u_0 = 2M/R$ and $v_0 = 4 \pi R^2 T^4$.

\subsection{Regular solutions}
Regular solutions to Eqs. (\ref{equ}) and (\ref{eqv}) have been studied in detail in Refs. \cite{PWZ, Cha1}. In a regular solution, the functions $u$ and $v$ vanish at the center. This means that the set of regular solutions  corresponds to the solution curve of the differential equation
\begin{eqnarray}
\frac{dv}{du} = \frac{2v (1 - 2 u - \frac{2}{3}v)}{(1 - u)(2v - u)}, \label{vu}
\end{eqnarray}
with initial condition $v(0) = 0$. This curve  is plotted in  Fig. 1. Each  point in this curve corresponds to boundary data $(u_0, v_0)$ that define a regular solution.

Fig. 1 shows that regular solutions are not defined for all values of $u_0$ or $v_0$. The point $P$    corresponds to the maximum value of $v_0$ allowed by a regular solution, and the point $Q$ to the maximum value of $u_0$ allowed by a regular solution. We denote the coordinates of $P$ as $(u_P, v_P)$ and those of $Q$ as $(u_Q, v_Q)$. Thus, for  regular solutions $u_0 \leq u_Q$ and $v_0 \leq v_P$. We find numerically that  $u_P \simeq 0.3861 $, $v_P \simeq 0.3416$,  $u_Q \simeq 0.4926$ and $v_Q \simeq 0.2463$.  In terms of the variables $(R, M, T)$, the inequalities above imply that for regular solutions,
 $TR^{1/2} \leq 0.406$ {\em and} $M/R \leq 0.246$. Hence, the restriction to the subset of regular solutions implies an upper bound to  both temperature $T$ and ADM mass $M$ for a given area $4\pi R^2$ of the bounding box.

\begin{figure}[tbp]
\includegraphics[height=6cm]{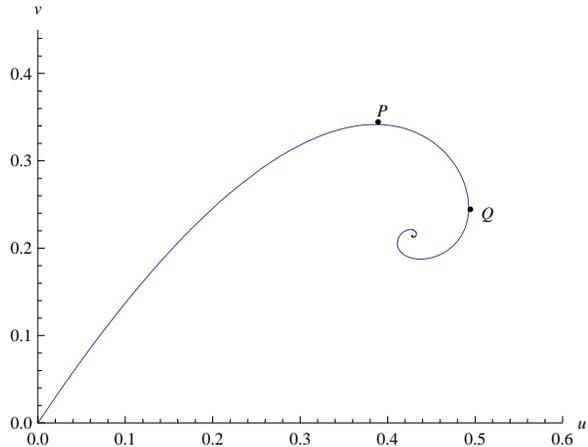} \caption{ \small The curve on the $u_0-v_0$ plane representing regular solutions to Eqs. (\ref{equ}--\ref{eqv}).  }
\end{figure}

Since the curve in Fig. 1 develops a spiral for $u_0 > u_P$, there are several different values of $v_0$ corresponding to each value of $u_0$, when $u_0 \in (u_P, u_Q)$. This is because the point $P$ in Fig. 1 signifies the onset of thermodynamic instability:
 $(\partial M/\partial T)_R$ becomes negative at  $P$  \cite{PL}.
However, restricting to the segment from the origin to the point $Q$, we can define a single-valued map $v_{reg}(u_0)$ that assigns a unique value $v_0 = v_{reg}(u_0)$ to each $u_0 \in [0, u_Q]$.

\subsection{Singular solutions}

\subsubsection{A `no-horizon' theorem}

In static spherically symmetric solutions to Einstein's equations an event horizon is identified by the condition $2m(r)/r = 1$, or equivalently $u(\xi) = 1$.
The singular solutions to Eqs. (\ref{equ}) and (\ref{eqv}) exhibit no event horizon, i.e., they are `naked'. The integration of the Oppenheimer-Volkoff equation from the boundary inwards encounters no horizons. This statement is a special case of a general theorem valid for a larger class of equations of state
   \cite{ST}, so the lack of horizons is not a special property of self-gravitating radiation.

  The theorem of Ref. \cite{ST} applies to solutions with continuous first derivatives of the energy density $\rho(r)$. However, it is conceivable that the energy density $\rho$ could be discontinuous or divergent on a horizon, a possibility that is not excluded by the theorem above.
   For this reason, we provide  an elementary proof that no horizon is encountered in the integration of  Eqs. (\ref{dm}--\ref{OV}), in which the  assumptions about the behavior of $\rho(r)$ on the horizon are weaker than those of Ref. \cite{ST}. The proof proceeds as follows.

\smallskip

 Assume that an horizon forms for a value  $\xi =  \xi_* \in (-\infty, 0)$, i.e.,  $u(\xi_*) = 1$. The horizon is reached by integration from the boundary inwards, and we are interested in the behavior of the solution as $\xi \rightarrow \xi_{*+}$. For $\xi > \xi_*$, $u < 1$, so we define $\epsilon(\xi) := 1 - u(\xi) > 0$. We assume continuity of first derivatives of $u$ and $v$  for all $\xi \in (-\infty, \xi_*)$, but make no such assumption for the behavior of $u(\xi) $ and $v(\xi)$ at $\xi = \xi_{*+}$
There are three possibilities for the behavior of $v(\xi)$ at the limit  $\xi \rightarrow \xi_{*+}$.
\begin{enumerate}

\item $lim_{\xi \rightarrow \xi_{*+}}v(\xi) = 0$. Then, in a sufficiently small neighborhood of $\xi_0$, Eq. (\ref{vu}) is approximated by
    \begin{eqnarray}
\frac{dv}{d\epsilon} = -\frac{2v}{\epsilon},
\end{eqnarray}
which is solved by $v = c \epsilon^{-2}$ for a constant $c$. On the horizon ($\epsilon = 0$) $v(\xi) \rightarrow \infty$, thus contradicting the initial assumption. The only exception is the trivial case $c = 0$, which corresponds to   $v(\xi) = 0 $ for all $\xi > \xi_*$.

\item $lim_{\xi \rightarrow \xi_{*+}}v(\xi) = \bar{v} >0$. In a sufficiently small neighborhood of $\xi_0$,  and assuming $\bar{v} \neq \frac{1}{2}$, Eq. (\ref{vu}) is approximated by
\begin{eqnarray}
\frac{dz}{d\epsilon} =  \frac{2 \bar{v} (1 + \frac{2}{3}\bar{v})}{\epsilon(2 \bar{v} - 1)},
\end{eqnarray}
where we wrote $z = v - \bar{v}$. The solution is $z = \ln \left(c \epsilon^{\frac{2 \bar{v} (1 + \frac{2}{3}\bar{v})}{2 \bar{v} - 1}}\right)$, for a constant $c>0$. Then, on the horizon  either $z \rightarrow \infty$ or $z \rightarrow -\infty$, thus contradicting the initial assumption. Similarly, one can show that the case $\bar{v} = \frac{1}{2}$ also leads to contradiction.

\item $lim_{\xi \rightarrow \xi_{*+}}v(\xi) = \infty$. In a sufficiently small neighborhood of $\xi_0$, Eq. (\ref{vu}) is approximated by
    \begin{eqnarray}
    \frac{dv}{d\epsilon} = \frac{2v}{3\epsilon},
    \end{eqnarray}
    with solution $v = c \epsilon^{2/3}$. At the horizon, $v \rightarrow 0$, thus contradicting the initial assumption.
\end{enumerate}
We conclude that no singularity forms from the integration of Eqs. (\ref{equ}--\ref{eqv}) from the boundary inwards, except for the trivial case that $v(\xi) = 0 $ for all $\xi > \xi_*$. This exception corresponds to a boundary condition $\rho(R) = 0$, or equivalently $T(R) = 0$. Given an non-zero value of the ADM mass $M$, this solution represents a box containing  a Schwarzschild horizon at $r_S = 2M$ and no radiation. The solution cannot be continued to  $r < r_S$, because the Killing field $\frac{\partial}{\partial t}$ is spacelike in the Schwarzschild interior.

A consequence of the no-horizon theorem above is that no solutions with $m(0) > 0$ exist. For, if such a solution exists, it  satisfies $\lim_{r \rightarrow 0} u = \infty$ as $r \rightarrow 0$. Since
 $u < 1$  at the boundary, continuity of $u$ implies that $u(r_*) = 1$ for some $r_* \in (0, R)$, thus contradicting the no-horizon theorem.

\smallskip

The lack of an horizon in the singular solutions is  essential for the  consistent thermodynamic description of the system. This can be seen from the following argument. Let the system be initially in a state corresponding to a regular solution.  We use a quasi-static, adiabatic process to compress the gas and bring it into a state corresponding to a value of $u_0$ for which no regular solution exists. We then reverse this process and return  the system to its initial state.  If the singular solution involves a horizon, the reversed process would lead to the horizon's disappearance, and thus to a violation of the second law of black hole mechanics. It is therefore significant that the only solutions with an horizon have vanishing temperature at the boundary. These states cannot be reached in a finite number of steps by an initial state with $T \neq 0 $.

\subsubsection{Properties of the singular solutions}

Numerical integration of Eqs. (\ref{equ}) and (\ref{eqv}) from the boundary inwards shows that there exist
 two qualitatively different types of singular solutions, which we denote as type I and type II. Their main features are the following.

 Type I solutions correspond to  small values of $v$ near the boundary, and $u$ increasing as we integrate from the boundary inwards. In the neighborhood of a point, $r_{c}$, $u$ blows up, approaching, but never reaching, the value $u = 1$ that defines an horizon. Integrating the solution to $r < r_c$, we find that $u$ decreases rapidly and tends to $- \infty$ as $r \rightarrow 0$. The energy density $\rho$   has a very sharp peak at a value of $r$ slightly smaller than $r_{c}$, and then decays to $0$ at $r = 0$. This implies that most  of the photon gas  is concentrated in a narrow shell around $r_c$. Since in this region $u$ is very close to unity,  these solutions describe a `wall' formed around a `deformed' horizon, similar to the brick wall configurations proposed by 't Hooft \cite{thooft}. In type II solutions, $u$ is an increasing function of $r$, and it tends to $-\infty$ at $r \rightarrow 0$. In these solutions,the density $\rho$ typically exhibits a maximum at intermediate values of $r$. Some properties of type II solutions are discussed in Ref. \cite{ZP}.

The two classes of solutions above are not exhaustive, and they are defined in terms of qualitative characteristics rather than precise mathematical conditions. Nonetheless, the distinction is significant. For   $u_0 < u_P$,
 type I solutions correspond to pairs $(u_0, v_0)$ such that $v_0 < v_{reg}(u_0)$ and type II solutions to pairs $(u_0, v_0)$ such that $u_0 > v_{reg}(u_0)$. For $u_P < u_0 < u_Q$),
  the interchange between the two types of solution is rather complex, as there exist several regular solutions for each value of $u_0$. For $u_0 > u_Q$ the qualitative distinctions between the two types of solution are gradually lost. In general, for $v_0/u_0 << 1$ the solutions are  of type I and for $v_0/u_0 >> 1$ they are of type II.

Fig. 2  plots  the typical behavior of the mass function $m(r)$ and the energy density $\rho(r)$ as functions of $r$ for the two types of singular solutions. The dimensionless variables $m(r)/M$, $R^3 \rho(r)/M$ and $r/R$ are employed in these plots.

\begin{figure}[tbp]
\includegraphics[height=5cm]{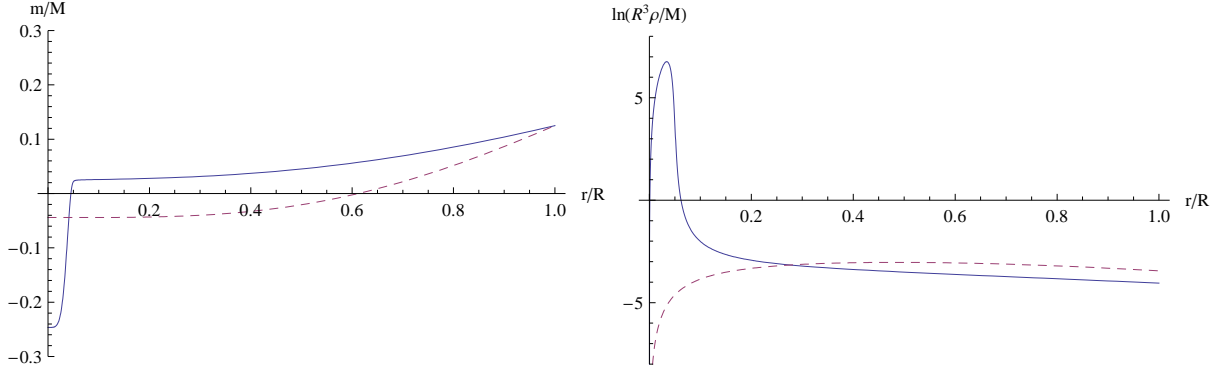} \caption{ \small The normalized mass function $m(r)/M$ as a function of $r/R$ and the normalized density $R^3\rho(r)/M$ as a function of $r/R$ for a type I solution (solid curve) and a type II solution (dashed curve). The sudden drop in $m(r)$ at $r\simeq 0.05 R$ in the  type I solution  coincides with the peak in the density. Note that the peak in the density is so sharp that we had to use a logarithmic scale in the vertical axis, in order to include it in the plot. The type I solution plotted here corresponds to $(u_0, v_0) = (0.25, 0.22)$ and the type II solution to $(u_0, v_0) = (0.25, 0.40)$. }
\end{figure}

The two types of solution do not differ in their approach to the singularity at $r = 0$: $m(0) = - M_0$ for some constant $M_0 > 0$. From  Eqs. (\ref{dm}---\ref{OV}), we find that for $r$ near $0$
\begin{eqnarray}
m(r) &=& - M_0 + \frac{1}{5}k r^5\\
\rho(r) &=&  \frac{k}{4\pi} r^2, \
\end{eqnarray}
for some constant $k$. It is convenient to employ the dimensionless parameters $\mu_0 = M_0/R$ and $\kappa = kR^4$, because these are functions of $u_0$ and $v_0$ alone. This can be seen from Eqs. (\ref{equ}--\ref{eqv}), whose solutions have an asymptotic behavior
\begin{eqnarray}
u = -2\mu_0e^{-\xi} \hspace{0.6cm} v = \kappa e^{4 \xi} \label{asympt}
\end{eqnarray}
as $\xi \rightarrow - \infty$.

The values of $M_0$ and $k$ are the same for all points $(R, M, T)$  of the same solution curve. This fact, together with the asymptotic behavior Eq. (\ref{asympt}), imply that $\mu_0$ and $\kappa$ satisfy the following equations.
\begin{eqnarray}
(2v_0 - u_0)\frac{\partial \mu_0}{\partial u_0} +  \frac{2v_0 (1 - 2 u_0 - \frac{2}{3}v_0)}{1-u_0} \frac{\partial \mu_0}{\partial v_0} &=& - \mu_0 \label{hom1}
\\
(2v_0 - u_0)\frac{\partial \kappa}{\partial u_0} +  \frac{2v_0 (1 - 2 u_0 - \frac{2}{3}v_0)}{1-u_0} \frac{\partial \kappa}{\partial v_0}&=& 4 \kappa. \label{hom2}
\end{eqnarray}

The  functions $\mu(u_0,v_0)$ and $\kappa(u_0, v_0)$ will be important for the construction of the singularity entropy in Sec. 3. Numerical evaluation shows that they possess the following properties.
\begin{enumerate}{}
\item For fixed $u_0$, $\mu_0(u_0, v_0) \sim v_0^{-1}$ and $\kappa(u_0, v_0) \sim v_0^{-1}$ as $v_0 \rightarrow 0$.
\item For fixed $u_0$, $\mu_0(u_0, v_0) \sim v_0^{3}$ and $\kappa(u_0, v_0) \sim v_0^{3}$ as $v_0 \rightarrow \infty$.
\item For fixed $u_0$, both $\mu_0$ and $\kappa_0$ diverge to infinity as $v_0$ approaches from below the value $v_{reg}(u_0)$ characterizing a regular solution.  $\kappa$ diverges more strongly: $\mu_0^2/\kappa \rightarrow 0$ as $v_0 \rightarrow v_{reg}(u_0)_-$.
     Both functions are continuous in their approach to $v_{reg}(u_0)$ from above. In particular, the function $\mu_0^2/\kappa$ is continuous at $v_{reg}(u_0)$ and is equal to zero. The characteristic behavior of $\mu_0$ and $\kappa$ in the vicinity of $v_{reg}(u_0)$ is illustrated in Fig. 3.
\end{enumerate}

\begin{figure}[tbp]
\includegraphics[height=5cm]{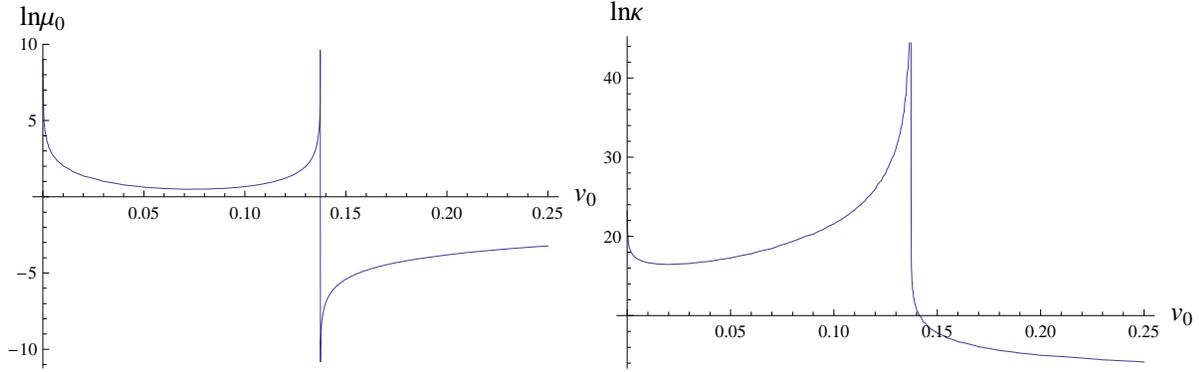} \caption{ \small Plots of $\ln \mu_0(u_0,v_0)$ and $\ln \kappa(u_0,v_0)$ for $u_0 = 0.1$ and different values of $v_0$ near the point $v_0 =v_{reg}(u_0) \simeq 0.1379$ corresponding to the regular solution.}
\end{figure}

In the vicinity of $r = 0$ the metric Eq. (\ref{ds2}) becomes
\begin{eqnarray}
ds^2 = - (1 - u_0) \sqrt{\frac{v_0}{\kappa}} \frac{R}{r} dt^2 + \frac{r dr^2}{2 \mu_0R} + r^2 (d\theta^2 + \sin^2\theta d \phi^2). \label{ds20}
\end{eqnarray}
The proper radius coordinate corresponding to the metric Eq. (\ref{ds20}) is $x = \frac{2}{3} r^{3/2}/\sqrt{2\mu_0R}$. A two-sphere of proper radius $x$ around $r = 0$ has area equal to $4 \pi (\frac{9}{2} \mu_0R)^{2/3} x^{4/3}$. This implies that the spacetime is not locally Minkowskian around $r = 0$ (since in that case the area should be $4 \pi x^2$), and that the point $r = 0$ is a conical singularity. The fact that $\rho \rightarrow 0$ as $r \rightarrow 0$ implies that the stress-energy tensor vanishes on the singularity, and, hence, unlike conical singularities in two spatial dimensions \cite{JH}, the conical singularities appearing in self-gravitating radiation cannot be interpreted as particles.

A singularity not hidden within an event horizon is usually thought of as  violating the cosmic censorship hypothesis (CCH) \cite{cch}, and thus it may be deemed unphysical. The main motivation for the CCH is to guarantee the preservation of  causality and predictability in general relativity, which are threatened by the existence of naked singularities. The greatest problem is posed by curvature singularities, such as the ones appearing in the black hole interior. However, the conical singularities appearing in self-gravitating radiation are rather `benign', in the sense that they do not give rise to inextensible causal geodesics.
Causal radial geodesics satisfy the equation
\begin{eqnarray}
\frac{1}{2} \dot{r}^2 = -g\frac{ \mu_0 R }{r} +\frac{\epsilon^2\mu_0R}{\alpha}, \label{geodesics}
\end{eqnarray}
where $\alpha = (1 - u_0) \sqrt{\frac{v_0}{\kappa}} R$, $\epsilon = \alpha \dot{t}/r$ is the energy corresponding to the geodesic, $g = 1$ for timelike geodesics and $g = 0$ for null geodesics. The derivative in Eq. (\ref{geodesics}) corresponds to proper time in timelike geodesics and to an affine parameter $\lambda$ in null geodesics.

From Eq. (\ref{geodesics}) it is evident that no timelike geodesics reach the singularity: incoming geodesics bounce back at a minimal radius $r_{min}= \alpha/\epsilon^2$. Null geodesics satisfy $\dot{r} = \pm \sqrt{\frac{\epsilon^2\mu_0R}{\alpha}}$ and they reach $r = 0$ within a finite value of their affine parameter (and also of the coordinate time $t$). An incoming null geodesic reaching the singularity simply becomes an outcoming null geodesic.

\section{Thermodynamic consistency and the entropy of singularities}

\subsection{The identification of singularity entropy}

As discussed in Sec. 2.1, self-gravitating radiation in a spherical box is  described by two independent thermodynamic variables, which can be chosen to be the ADM mass $M$ and the `radius' $R$. However, the structure equations (\ref{dm}) and (\ref{OV}) require the specification of {\em three} independent variables.

In usual thermodynamic systems the values of any unconstrained parameters are determined by the maximum-entropy principle, namely, the statement that their equilibrium values maximize the entropy for a given value of the energy \cite{Callen}. In order to implement this principle, it is necessary to construct an entropy functional that depends on the three independent variables required for the specification of a unique solution of the structure equations (\ref{equ}) and (\ref{eqv}). Thus, we must construct an entropy functional $S(M, T, R)$ or equivalently $S(u_0, v_0, R)$.

To this end, we first consider the total entropy $S_{rad}$ of the radiation in the box. The entropy density $s$ of radiation is $s = \frac{4}{3} \rho^{3/4}$. Hence,
\begin{eqnarray}
S_{rad} = \frac{16 \pi}{3} \int_0^R dr \frac{r^2 \rho^{3/4}}{\sqrt{1 - \frac{2m}{r}}} = \frac{4}{3} (4 \pi)^{1/4} \int_0^R \frac{r^{1/2} v^{3/4}}{\sqrt{1-u}}. \label{srad}
\end{eqnarray}

For solutions to Eqs. (\ref{equ}--\ref{eqv}), the integrand in Eq.(\ref{srad}) is a total derivative \cite{PWZ}, i.e.,
\begin{eqnarray}
\frac{r^{1/2} v^{3/4}}{\sqrt{1-u}} = \frac{d}{dr} \left( \frac{v + \frac{3}{2}u} {6v^{1/4}\sqrt{1-u}} r^{3/2}             \right).
\end{eqnarray}
Hence, the radiation entropy is a sum of two terms
\begin{eqnarray}
S_{rad}(u_0, v_0,R) = S_1(u_0, v_0, R) + S_0(u_0, v_0, R), \label{srad2}
\end{eqnarray}
where $S_1$ is defined by the field values at the boundary $r = R$
\begin{eqnarray}
S_1(u_0, v_0, R) = \frac{2}{9} (4 \pi)^{1/4} \frac{v_0 + \frac{3}{2}u_0} {v_0^{1/4}\sqrt{1-u_0}} R^{3/2}
\end{eqnarray}
and $S_0$ by the field values at $r = 0$
\begin{eqnarray}
S_0(u_0, v_0, R) = - \frac{2}{9} (4 \pi)^{1/4} \lim_{r\rightarrow 0} \left( \frac{v + \frac{3}{2}u} {v^{1/4}\sqrt{1-u}} r^{3/2}\right) = \frac{1}{3} (4 \pi)^{1/4} \left(\frac{4\mu_0(u_0, v_0)^2}{\kappa(u_0, v_0)}\right)^{1/4} R^{3/2}, \label{s0}
\end{eqnarray}
where the limit was taken using Eq. (\ref{asympt}) for the behavior of the functions $u$ and $v$ as $r \rightarrow 0$ ($\xi \rightarrow - \infty$). We note that
the term $S_0$ vanishes for regular solutions.

\medskip

The usual study of  static spherically symmetric solutions of Einstein's equations invokes the condition that the spacetime metric is regular at the center. The regularity condition is motivated by the perception of singularities as pathologies in the description of space-time geometry that should be avoided in order to guarantee the causality and predictability of the theory. However, as shown in Sec. 2.3.2,  this is not a problem for the singular solutions considered in this paper. In Sec. 3.2, we will also argue that the consideration of singular solutions does not necessitate the acceptance of singularities as genuine spacetime points, at least in the context of the approach developed in this paper.
However, the {\em a priori} exclusion of singular solutions as unphysical, implies that there can be no description of the system in the sector of the state space where regular solutions do not exist. There will be total unpredictability for any thermodynamic operation that takes the system into this sector.

Moreover, the lack of a more fundamental justification for the regularity condition is conceptually problematic. If  general relativity arises as the macroscopic limit of an underlying microscopic theory, there should be some physical mechanism guaranteeing the stability of regular solutions against statistical or quantum fluctuations of the underlying structure. At the macroscopic level the only language we possess to describe such mechanisms is that of thermodynamics.

There is therefore good reason for providing a thermodynamic interpretation of the regularity condition. This implies that one should make no {\em a priori} distinction between regular and singular solutions. The regular solutions should be identified as the physical states of the system by the fundamental condition of thermodynamic equilibrium, namely, the principle of maximum entropy. This means that regular solutions should correspond to  global maxima of the entropy functional, for any given values of  energy $M$ and  box area $4 \pi R^2$.

\begin{figure}[tbp]
\includegraphics[height=7cm]{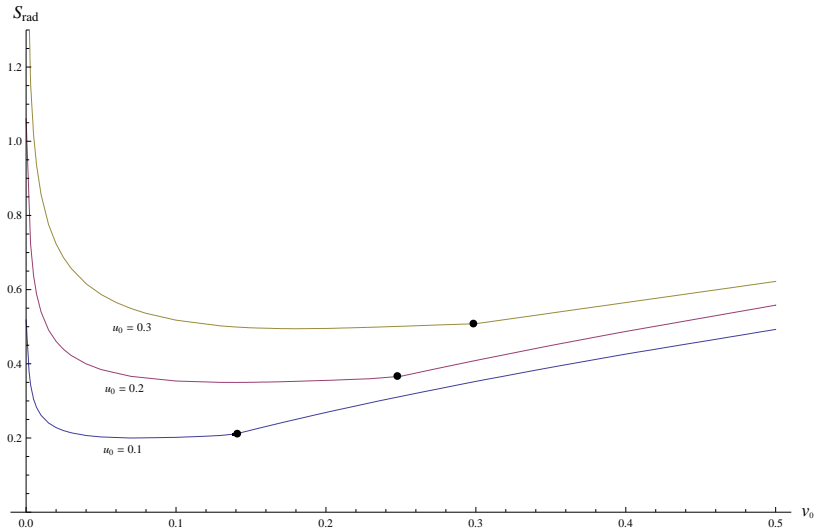} \caption{ \small $S_{rad}/R^{3/2}$ as a function of $v_0$ for  different values of $u_0$. The radiation entropy does not have a maximal value, it tends to infinity at the limits $v_0 \rightarrow 0$ and $v_0 \rightarrow \infty$. The dots denote the regular solutions. }
\end{figure}

However, the radiation entropy $S_{rad}$ of Eq. (\ref{srad2}) does not allow for the implementation of the maximum-entropy principle. $S_{rad}$ has no global maximum  for any fixed values of $R$ and $M$.
The typical dependence of $S_{rad}$ on $v_0$ for fixed values of $u_0 = 2M/R$ and $R$ is depicted in Fig. 4. There, we see that $S_{rad} \rightarrow \infty$ as $v_0 \rightarrow 0$ and as $v_0 \rightarrow \infty$. In fact, the regular solution is close to the global {\em minimum} of the entropy functional.
 This is quite problematic, because  $S_{rad}$ has the same behavior even at the limit of
 near-vanishing gravitational interaction ($R>>1$ and  $M/R$ arbitrarily close to zero). This would imply that   the ordinary phase of thermal radiation is thermodynamically unstable even in presence of a tiny gravitational self-interaction.

 The behavior of $S_{rad}$, described above, is not unexpected: it is an analogue of the well known gravothermal catastrophe \cite{gth}. However, it means that the thermodynamic justification of the regularity condition cannot be obtained by the consideration of the radiation entropy alone. The only solution to this problem  is to assume that the entropy functional also includes a contribution from the gravitational field.
 In static spherical symmetric configurations, there are no local gravitational degrees of freedom, so one expects that any gravitational contribution to entropy is topological, i.e., it corresponds to the properties of the singularities (or, possibly, to the properties of an interior boundary). If the entropy of the singularities is significantly lower than the entropy of locally Minkowskian space, then the regularity condition could follow as a consequence of the maximization of entropy.

 A  different line of thought suggesting that singularities in general relativity could be associated to entropy originates from Penrose
 \cite{Pen}. Penrose's idea is that different types of singularity are distinguished by their entropic content. In particular, strong curvature singularities such as the one in the Schwarzschild-Kruskal spacetime correspond to high entropy, while `mild' singularities (for example, the Big-Bang where the spacetime is conformally regular) correspond to low entropy. The primary aim of this proposal is to provide a consistent thermodynamic description of cosmology. However, the idea applies naturally to the present system. Loosely speaking,  the singularities of self-gravitating radiation are mild (no geodesic incompleteness), so Penrose's conjecture would suggest that they correspond to low entropy, as would be required for a thermodynamic explanation of the regularity conditions.

\medskip

Thus, the requirement of a consistent thermodynamic interpretation for the regularity condition leads to the conjecture that the total entropy of the system can is a sum of the radiation entropy $S_{rad}$ and a term $S_{sing}$ for the entropy of singularities.
\begin{eqnarray}
S_{tot} = S_{rad} + S_{sing}. \label{stotdef}
\end{eqnarray}

 Regular solutions  have maximum entropy, only if  the entropy of the singularities is lower than the entropy of locally Minkowskian geometry.  Otherwise, a locally Minkowskian spacetime would  be unstable even in   absence of matter. It is convenient to  choose $S_{tot}$  so that the solutions with $M = 0$ have zero entropy for all $R$. The term $S_{sing}$ is then negative-valued. It turns out that with this convention, the total entropy $S_{tot}$ may take negative values. This contradicts the physical requirement that entropy is positive-valued. To make $S_{tot}$ positive-valued, we must add a reference term $S_{flat}(R)$, corresponding to the gravitational entropy of flat spacetime within a spherical box of radius $R$. However,
 $S_{flat}(R)$ does not depend on $u_0$ and $v_0$, and hence, it does not affect  the specification of the equilibrium states through the maximum-entropy principle. For this reason,  $S_{flat}(R)$ is ignored in the present context. The entropy of flat  spacetime, viewed as a reference state for the system, cannot be fixed by purely thermodynamic arguments; it requires the knowledge of the microstates in the underlying theory.   In Sec. 3.3, we show that the existence of a lower bound in $S_{tot}$ also  implies a lower bound for $S_{flat}(R)$.

   We point out that a non-zero value of $S_{flat}(R)$ does not contradict the idea that Minkowski spacetime is the unique lowest energy state of gravity, and thus, that it should be assigned zero entropy. $S_{flat}(R)$ refers to the entropy of the degrees of freedom {\em localized within the box of radius $R$}. Even if the entropy of a globally Minkowski spacetime is zero, a microscopic definition of $S_{flat}(R)$ would require tracing out the degrees of freedom outside the box. Hence, even if  globally Minkowski spacetime is the vacuum in a quantum theory of gravity, the reduced state within the box is expected to be mixed due to correlations across the boundary, and thus possessing non-zero entropy. An analogous calculation for a scalar field leads to an entropy proportional to the area of the boundary \cite{Sred}.

Lacking a general theory for gravitational entropy, we can construct $S_{sing}$ only by constraining it with requirements of physical and mathematical consistency. These requirements are the following:
\begin{enumerate}
\item The application of the maximum-entropy principle to general self-gravitating systems ought to recover the  successful theory of stellar stability, which is based on the regularity condition. This means that the regular solutions must correspond to global maxima of the entropy $S_{tot}$. For self-gravitating radiation in particular, we should recover a large segment of the $u_0-v_0$ curve of Fig. 1; at the very least the segment starting from the origin and ending at the point $P$.

\item The radiation entropy $S_{rad}$ has the following behavior under rescalings: $S_{rad}(u_0, v_0, \lambda R) = \lambda^{3/2} S_{rad}(u_0, v_0, R)$, for any $\lambda > 0$. If $S_{sing}$ has a different behavior under rescalings, then the value of $v_0$ for maximum-entropy configurations would carry a dependence on $R$ in addition to the dependence on $u_0$, and we would not be able to recover  the regular solutions as global maxima of entropy. This means that  the singularity entropy  should also rescale as $S_{sing}(u_0, v_0, \lambda R) = \lambda^{3/2} S_{sing}(u_0, v_0, R)$.

    \item The existence of global maxima of entropy requires that   $S_{tot}(u_0, v_0, R)$ is bounded from above, for fixed values of $u_0$ and $R$. This implies, in particular, that $S_{tot}$ converges to finite values at the limits $v_0 \rightarrow 0$ and $v_0 \rightarrow \infty$, and that neither of these values is a global maximum. For constant $R$, $S_{tot}$ should also be bounded from below, so that $S_{tot}$ can be made positive-valued by the addition of a term $S_{flat}(R)$ that has no dependence on $u_0$ and $v_0$.

        \item The term $S_{sing}$ depends only on the properties of the singularity, i.e., on (limits of) values of the field variables at $r=0$. Assuming that $S_{sing}$ depends only on the values of $m(r)$ and of its first derivative at the limit $r \rightarrow 0$, $S_{sing}$ is a functional of $M_0$ and $k$ alone.
\end{enumerate}

The four conditions above are very restrictive and the existence of a functional $S_{sing}$ that satisfies them is highly non-trivial. To find $S_{sing}$, we observe that $S_{rad} \sim v_0^{-1/4}$ as $v_0 \rightarrow 0$ and $S_{rad} \sim v_0^{3/4}$ as $v_0 \rightarrow \infty$.  $S_{sing}$ should have the same asymptotic behavior with $S_{rad}$   if property 3 is to be satisfied. The only combination of $\mu_0$ and $\kappa$  with this asymptotic behavior at the limit $r \rightarrow 0$, which vanishes in the absence of singularities, is $(\mu_0^2/\kappa)^{1/4}$.

Thus, given Eq. (\ref{s0}), the natural candidate for the singularity entropy   is a term
 $S_{sing} = b S_0$ for some constant $b$. We determine $b$ by the asymptotic behavior of $S_{rad}/S_0$ at the limits $v_0 \rightarrow 0$ and $v  \rightarrow \infty$. Since two conditions are   employed in order to determine a single parameter, the generic expectation is that the problem admits no solution. It is then quite remarkable that the two limits identify the same value of $b$. We find that $\lim_{v_0 \rightarrow 0} S_{rad}/S_0  = \lim_{v_0 \rightarrow \infty} S_{rad}/S_0 = 6$. Thus, $b = -6$ is the only value compatible with the required asymptotic behavior of $S_{sing}$. It follows that
\begin{eqnarray}
S_{tot} = S_{rad} - 6 S_0 = S_1 - 5 S_0. \label{stot}
\end{eqnarray}

The entropy $S_{tot}$ Eq. (\ref{stot})  satisfies conditions 2-4 above.
The issue is whether $S_{tot}$ also satisfies condition 1, i.e., whether its global maxima correspond to regular solutions. Impressively, it does so. $S_{tot}/R^{3/2}$ as a function of $v_0$ for different values of $u_0$ is plotted in Fig. 5. For each $u_0$, there is a clear maximum at a specific value of $v_0$, which we denote as $v_{eq}(u_0)$. For $u_0 < u_Q$, the maxima of $S_{tot}$ coincide with regular solutions.
For $v_0 < v_{eq}(u_0)$ the singular solutions are of type I, while for $v_0 > v_{eq}(u_0)$ they are of type II. The dependence of $S_{tot}$ on $v_0$ is different in the two types of solution. This results to  non-smooth behavior of $S_{tot}$ at the maxima: the first derivative of $S_{tot}$ is discontinuous at $v_0 = v_{eq}(u_0)$.

 The behavior of $S_{tot}$ changes for $u_0 > u_P$, as can be inferred from the presence of the spiral in Fig. 1. $S_{tot}$ develops additional local maxima, reflecting the presence of multiple regular solutions with same value of $u_0$. These solutions are not maxima with respect to the radiation entropy $S_{rad}$. The derivative of $S_{tot}$ is discontinuous at the local entropy maxima. The local maximum with the largest value of $v_0$ is also the global maximum, and corresponds to the equilibrium configuration. The local maxima correspond to metastable configurations.

 In general, metastable configurations in self-gravitating systems may be characterized by very long decay times \cite{Cha2}, and thus considered as physical. An estimation of the relevant timescales is not possible in the present context, because our analysis lies entirely within equilibrium thermodynamics; hence, we cannot make any statement about the physical relevance of the metastable solutions.
 However, we note that as $u_0$ increases, the local maxima approach each other and at  $u_0 = u_Q$ they merge into a single maximum and the derivative of $S_{tot}$ becomes everywhere continuous.
 The existence of a single maximum persists for all values $u_0 > u_Q$. The fact that the distance between the local maxima becomes arbitrarily small near the critical point $u = u_Q$ strongly suggests that the local maxima would strongly contribute in the properties of the phase transition in a statistical-mechanics description of this system.

We also note that there is a large number of regular solutions for  values of $u_0$ near the center of the spiral in Fig. 1. We have verified that these regular solutions indeed correspond to local entropy maxima, at least within the degree of accuracy allowed by our numerical evaluation.

Fig. 5 shows $S_{tot}$ for a small range of values of $v_0$ around the local maxima. However, an equilibrium solution corresponds to a {\em global} maximum of $S_{tot}$. In Fig. 6, we plot $S_{tot}$ for a larger range of values of $v_0$, capturing also its asymptotic behavior for small and large values of $v_0$, and showing that the local entropy maxima are also global. Fig. 6 also illustrates the fact that the entropy $S_{tot}$ is bounded from below.

\begin{figure}[tbp]
\includegraphics[height=8cm]{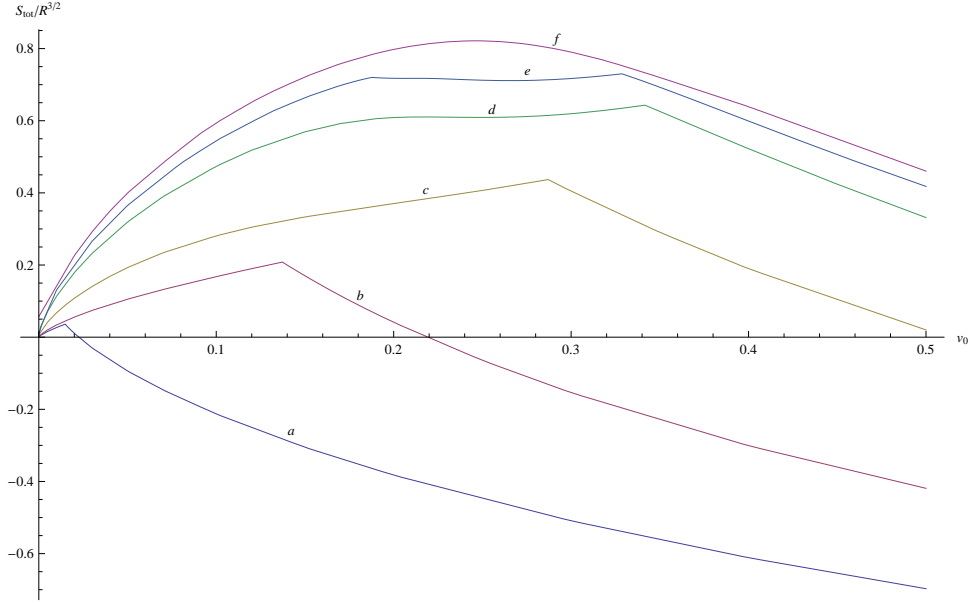} \caption{ \small  $S_{tot}(u_0, v_0, R)/R^{3/2}$ as a function of $v_0$ for different values of $u_0$. Curve a corresponds to $u_0=0.01$, curve b to $u_0 = 0.1$, curve c to $u_0 = 0.25$, curve d to $u_0 = u_P \simeq 0.3861$, curve e to $u_0 = 0.44$ and curve f to $u_0 =u_Q \simeq 0.4923$. The maxima of curves d and f identify the points P and Q of Fig. 1, respectively.
For  $ u_P < u_0 < u_Q$ the curves have two, or more, local maxima, which correspond to different regular solutions with the same value of $u_0$.  The maximum with the larger value of $v_0$ is the global maximum.}
\end{figure}

\begin{figure}[tbp]
\includegraphics[height=9.5cm]{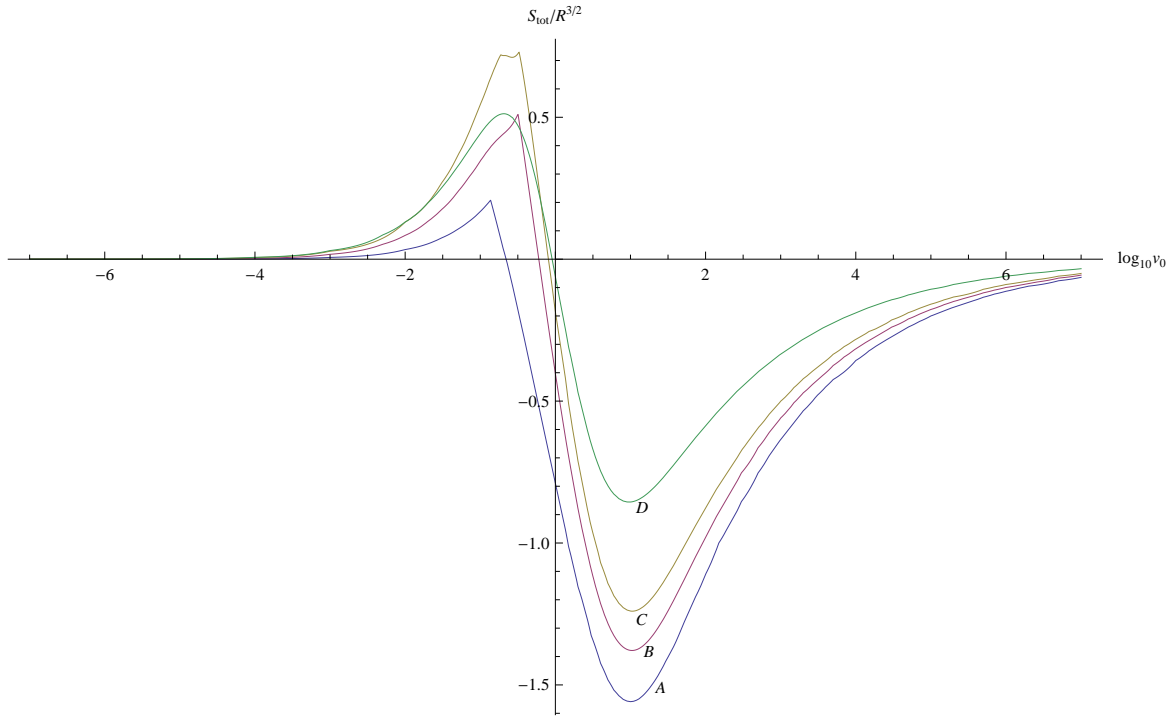} \caption{ \small  $S_{tot}(u_0, v_0, R)/R^{3/2}$ as a function of $\log_{10} v_0$ for different values of $u_0$. Curve $A$ corresponds to $u_0 = 0.1$, curve $B$ to $u_0 = 0.3$, curve $C$ to $u_0 = 0.44$ and curve $D$ to $u_0 = 0.75$.}
\end{figure}

The question arises whether the choice $S_{sing} = - 6 S_0$ for the entropy of the singularities is the unique expression satisfying conditions 1-4 above. In general, if the four conditions admit one solution $S_{sing}$, then any function $f(S_{sing})$ such that (i) $f(S_{sing}) \rightarrow S_{sing}$ at the limits $v_0 \rightarrow 0 $ and $v_0 \rightarrow \infty$, and (ii) has the same  global maxima with $S_{sing}$,  provides an alternative solution. In principle, there may exist many such functions, even though their explicit construction would be rather artificial.

However, a physically natural assumption is that
  $S_{sing}$ is a boundary term obtained from the integration of a total divergence $\partial_r \cal{S}$, where  ${\cal S}$ a functional of the fields $m(r)$ and $\rho(r)$. It is also natural to expect that $S_{sing}$  depends only on the topological properties of the singularity which, given the scaling symmetry of the system, are the same in all geometries that correspond to the same solution curve of Eqs.  (\ref{equ}) and (\ref{eqv}). This property is also satisfied by $\mu_0$ and $\kappa$.
  Then, $S_{sing}$ should satisfy an equation of the same form as  Eqs. (\ref{hom1}--\ref{hom2}), namely,
\begin{eqnarray}
 (2v_0 - u_0)\frac{\partial S_{sing}}{\partial u_0} +  \frac{2v (1 - 2 u_0 - \frac{2}{3}v_0)}{1-u_0} \frac{\partial S_{sing}}{\partial v_0} =  \gamma S_{sing}, \label{hom3}
\end{eqnarray}
where $\gamma$ is a constant defined by the asymptotic behavior of  ${\cal S}$ as $r \rightarrow 0$: ${\cal S} \sim r^{\gamma}$. The choice $S_{sing} = -6 S_0$ is the only one that satisfies Eq. (\ref{hom3}) in addition to the conditions (i) and (ii) above, for $\gamma = - \frac{3}{2}$.

While we believe that the above argument for the uniqueness of $S_{sing} = - 6 S_0$ is quite plausible on physical grounds, the additional assumptions go beyond basic thermodynamic principles, and they may not provide a unique result when applied to other self-gravitating systems. It is necessary to perform a similar analysis on other systems before we can make a definite statement on this issue.

In any case, the choice $S_{sing} = -  6 S_0$ is the simplest possible expression for the singularity entropy. The existence of a simple solution to a strongly constrained problem is an indication that the singularity entropy is not an artificial construct  and that it represents a genuine feature of gravitational thermodynamics, that is fundamentally embedded into the structure of Einstein's equations.

\subsection{The physical interpretation of singularity entropy}

It is important to emphasize that the assignment of entropy to the singularities at $r = 0$ does not necessarily imply an interpretation of these singularities as actual spacetime points. Such an assumption does not enter into the arguments employed for the identification of $S_{sing}$. In particular, our results do not require a commitment to the physical existence of naked singularities (however benign those may be), or  to discount the possibility that quantum effects might smoothen out the singularity at the Planck scale. The key point here is that the term $S_0$ in Eq. (\ref{s0}) (and consequently $S_{sing}$) is associated to an {\em interior boundary} of the singular solutions to the Oppenheimer-Volkoff equations, and the application of the maximum entropy principle is well-defined whether the singularity corresponds to a spacetime point, or not.

 In order to explain this point, we consider that the interior solutions to the Oppenheimer-Volkoff equations are defined for values of $r \in [\epsilon, R]$, where $\epsilon$ is a   length cut-off. For example, $\epsilon$ may correspond to a scale where quantum or statistical effects beyond relativity become important. Let us denote the corresponding spacetime manifold as $M_{\epsilon}$. This can be embedded isometrically into a larger spacetime manifold $M$, which extends into the region $r < \epsilon$. For example, $M$ may be the union of the spherically symmetric solution $M_{\epsilon}$ with a homogeneous solution with suitable junction conditions on the surface $r = \epsilon$---similar to a construction in Ref. \cite{PWZ}. The continued solution needs not contain any spacetime singularities. It is presumably determined by a microscopic theory of gravity, and its precise form is irrelevant to the arguments in this paper.

  The reason is that the maximum entropy principle applies, irrespective of the nature of the spacetime extension. The radiation entropy $S_{rad}$ is a sum of the term $S_1$ defined on the exterior boundary $r = R$ and of a term $S_{\epsilon}$ defined on the interior boundary $r = \epsilon$. Since $\epsilon$ can be taken arbitrarily close to zero, $S_{\epsilon}$ is well approximated by $S_0$. In order to apply the maximum entropy principle, we consider a total entropy $S_{tot} = S_{rad} + S_{bound}$, where the term $S_{bound}$ is defined on the boundary $r = \epsilon$ and corresponds to the total entropy associated to the region  $r < \epsilon$.  The entropy $S_{bound}$ is specified by the requirement that the maximum entropy principle holds. That is, we seek an expression for $S_{bound}$, such that $S_{tot}$ is maximized for fixed $M$ and $R$. This is a well-defined maximization problem. Its solution
   is independent of the form of the extension in the region $r < \epsilon$. Whatever the physical configuration   in the region $r < \epsilon$ might be, its entropy equals  the value $S_{bound}$ that maximizes $S_{tot}$.

 The solution to the maximization problem depends only on the functional form of $S_{rad}$. Taking $\epsilon$ arbitrarily close to $0$, continuity implies that $S_{rad}$ is well approximated by the expression (\ref{srad2}), which depends on the fields' values at the singularity $r = 0$. It follows that at the limit $\epsilon \rightarrow 0$, $S_{bound}$ coincides with $S_{sing} = - 6 S_0$, as shown in Sec. 3.1. Thus, the limiting behavior of the fields near the singularity $r = 0 $ specifies the entropy $S_{sing}$, even if the singularity is not viewed as part of the spacetime.
  At present, the physical origin of $S_{sing}$ can only be a matter of conjecture: we can only identify it as a contribution to the entropy, defined on internal boundaries of solutions to Einstein' s equation, that is necessary for the existence of a consistent thermodynamic description.

\subsection{The implementation of the maximum-entropy principle}

The maximum-entropy principle asserts that  equilibrium configurations correspond to the maximum value of entropy for a given value of the {\em internal energy}. If we identify the internal energy with the ADM mass $M$, namely, the   energy measured by  observers at asymptotic infinity, then the maximum-entropy principle applied to the entropy function $S_{tot}$ Eq. (\ref{stot}) clearly implies that the regular solutions are the equilibrium configurations.

The ADM mass $M$ is a directly observable quantity for an observer outside the box. However, its identification with the thermodynamic internal energy is not mandatory. Even in non-gravitating systems, the internal energy does not always coincide with the total energy. Total energy depends on the reference frame: for equilibrium systems one usually identifies the internal energy with the total energy at the rest frame of the system's center of mass. In non-equilibrium systems, the   internal energy is obtained from the total energy by subtracting the contribution of energy related to the dynamics of the flow \cite{GM}.

A key problem in the identification of the ADM mass $M$ with the internal energy is that the singular  spacetimes considered here possess an internal boundary at the singularity $r = 0$. The mass function $m(r)$  starts from a negative value $-M_0$ at $r = 0$. Hence, one would expect that the value  $m(R) = M$ at the bounding box underestimates the internal energy of the system.

Since we deal with static solutions to Einstein's equations, a better candidate for the internal energy $U$ is the Komar integral,
\begin{eqnarray}
U = -\frac{1}{8 \pi} \int_{\partial \Sigma} dS_{\mu \nu} \sqrt{-g}\partial^{\mu} \xi^{\nu},
\end{eqnarray}
where the integration is over all boundaries of a Cauchy surface $\Sigma$, $dS_{\mu \nu}$ is the area two-form on the boundary and $\xi := \frac{\partial}{\partial t}$. A general spherically symmetric metric is of the form
\begin{eqnarray}
ds^2 = - f(r) dt^2 + \frac{dr^2}{1 - \frac{2m(r)}{r}} + r^2 (d\theta^2 + \sin^2\theta d \phi^2),
\end{eqnarray}
where positive-valued function $f(r)$ and $m(r)$ the mass function. Evaluating the Komar integral for this class of metrics, we obtain
\begin{eqnarray}
U = \left[\frac{r^2}{2} \sqrt{1 - \frac{2 m}{r}} \frac{f'}{\sqrt{f}}\right]_0^{\infty}. \label{komar}
\end{eqnarray}
At $r \rightarrow \infty$ (outside the box), the metric is a Schwarzschild solution with ADM mass $M$. For  $r$ close to $0$, the metric is given by Eq.(\ref{ds20}). Hence,  $f(r) = (1 - u_0) \sqrt{\frac{v_0}{\kappa}} \frac{R}{r}$ as $r \rightarrow 0$. Evaluating Eq. (\ref{komar}), we obtain

\begin{eqnarray}
U = M + \frac{1}{2} R \left(\frac{4 \mu_0^2}{\kappa}\right)^{1/4} v_0^{1/4} \sqrt{1 - u_0} = M + \frac{3}{2} S_0 T_{\infty}, \label{uuu}
\end{eqnarray}
where $T_{\infty} = T \sqrt{1 - u_0}$ is the temperature as measured by an observer at the asymptotic infinity and $S_0$ is given by Eq. (\ref{s0}).

The  maximum-entropy principle then implies that the equilibrium configurations are determined by the maximization of the total entropy $S_{tot}$ for fixed values of the radius $R$ and the Komar mass $U$. The scaling symmetry of $S_{tot}$ implies that the quantity
\begin{eqnarray}
\sigma(u_0, v_0) = S_{tot}(u_0, v_0, R)/ R^{3/2}, \label{sig}
\end{eqnarray}
does not depend on $R$. It is convenient to define the variable
\begin{eqnarray}
w := 2 U/R = u_0 + \frac{3}{5} \left( \frac{v_0}{4\pi}\right)^{1/4} \sigma_0 \sqrt{1- u_0}, \label{wdef}
\end{eqnarray}
where $\sigma_0 = S_{sing}/R^{3/2}$ depends only on $u_0$ and $v_0$.

Then, the equilibrium configurations correspond to pairs $(u_0, v_0)$ that maximize $\sigma(u_0,v_0)$, subject to the constraint that $w$ is constant. We denote the entropy-maximizing values of $u_0$ and $v_0$ at fixed $w$ as $u_{eq}(w)$ and $v_{eq}(w)$ respectively.
 The entropy of the equilibrium states is
\begin{eqnarray}
S_{eq}(U, R) = R^{3/2} \sigma[u_{eq}(2U/R), v_{eq}(2U/R)]. \label{seq}
\end{eqnarray}

For values of $u_0$ such that regular solutions exist ($u_0  \in[0, u_Q]$),  entropy maximization at constant $w_0$ gives the same result as entropy maximization at constant $u_0$. The proof of this statement is the following.

\smallskip

   We define the function $\sigma_{max}(u_0) := \sup_{v_0} \sigma(u_0, v_0)$, corresponding to the maximum of $\sigma$ at fixed values of $u_0$. As shown in Sec. 3.1, $\sigma_{max}$ is an increasing function of $u_0$, for $u_0 \in [0, u_Q]$. Then, $\sigma_{max}(u_0) \geq \sigma_{max}(u_0')$ for any $u_0' \leq u_0$.

Next, we consider a curve  of constant $w$. Eq. (\ref{wdef}) implies that for any point $(u_0, v_0)$ in this curve, $u_0 \leq w$. Hence, for all points $(u_0, v_0)$ in the curve of constant $w$
      \begin{eqnarray}
      \sigma(u_0,v_0) \leq \sigma_{max}(u_0) \leq \sigma_{max}(w). \label{ssw}
      \end{eqnarray}
For $w \in [0, u_Q] $, the maxima at constant $u_0$ correspond to regular solutions, and equality in Eq. (\ref{ssw}) is achieved for the regular solution that maximizes $\sigma$ at fixed value of $u_0 = w$. Thus, for $w \in [0, u_Q] $, the maxima of $\sigma$ at constant $u_0$ and the maxima of $\sigma$ at constant $w$ coincide.

\smallskip

We conclude that
 entropy maximization at constant internal energy $U$ and radius $R$
        identifies the regular solutions as the equilibrium configurations. We note that the proof above does not require the explicit form of $U$, only the fact that $U \geq M$.

 One might inquire whether the identification of the Komar mass with the internal energy $U$ removes the need for the introduction of the term $S_{sing}$ for the entropy of the singularities. Namely, whether the radiation entropy $S_{rad}$ has a maximum at the regular solutions for fixed values of $U$. The answer is negative:  regular solutions correspond to {\em local} maxima of $S_{rad}$ at fixed $U$, {\em but not to global maxima}. Numerical evaluation shows that  $S_{rad}\rightarrow \infty$ as $v_0 \rightarrow 0$, for any fixed value of $U$. The correct global behavior of the total entropy can only be guaranteed by the addition of the term $S_{sing}$ for the entropy of the singularities.

To summarize, we argued that the internal energy of the singular solution should be identified with the Komar mass Eq. (\ref{uuu}) rather than the ADM mass $M$, and that the maxima of the total entropy $S_{tot}$ at fixed values of internal energy $U$ and radius $R$ correspond to regular solutions. Next, we proceed to identify the entropy-maximizing configurations in the sector of the thermodynamic state space where regular solutions do not exist.

\subsection{Equilibrium singular solutions}

In Fig. 7, we plot the functions $v_{eq}(w)$ and $u_{eq}(w)$ that correspond to the maxima of $\sigma$ at fixed values of $w$. These functions determine the temperature $T$ at the bounding box and the ADM mass of the {\em equilibrium} configurations for fixed internal energy $U$ and box area $4 \pi R^2$, as follows.
\begin{eqnarray}
T(U, R) &=& \left[\frac{v_{eq}(2U/R)}{4 \pi R^2}\right]^{1/4} \\
M(U, R) &=& \frac{1}{2} R u_{eq}(2U/R).
\end{eqnarray}
  We note that the equilibrium solutions  span the whole range of parameters $U$ and $R$.  Moreover, there exists a unique solution for each pair $(M, R)$ and at least one solution for each pair $(T, R)$.

\begin{figure}[tbp]
\includegraphics[height=5cm]{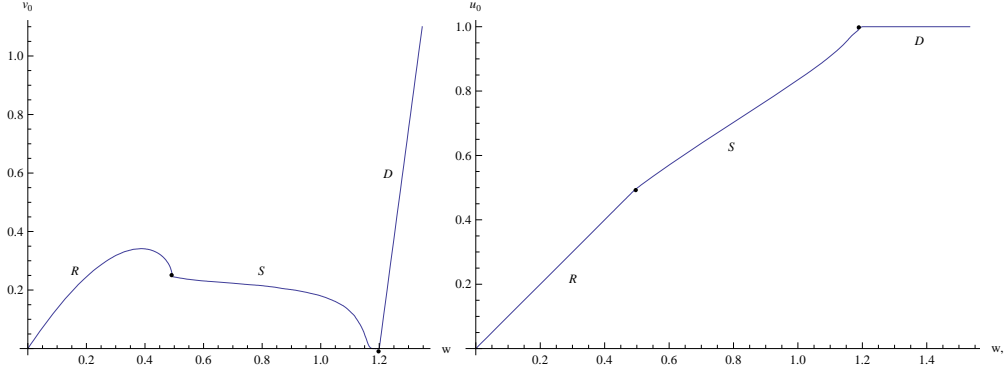} \caption{ \small  The curves $v_0 = v_{eq}(w)$ (left) and $u_0=u_{eq}(w)$ (right) corresponding to the equilibrium solutions.}
\end{figure}

As shown in Fig. 7, there exist  three different `phases' of equilibrium solutions.

 The first phase (R) corresponds to regular solutions for which $u_{eq}(w) = w$. It coincides with to the segment from the origin to the point Q in the curve of Fig.1.

 The second phase (S) corresponds to singular solutions with $u_0 \in (u_Q, 1)$ which maximize the entropy $S_{tot}$ for   $w \in [u_Q, 1.2)$. It is interesting to note that these solutions also maximize $S_{tot}$ for fixed $u_0$. Hence, the same set of equilibrium solutions are obtained even by identifying the internal energy with the ADM mass $M$. The S-phase consists  of type II solutions , except for values of $u_0$ near $1$, where  the qualitative distinction between types I and II is not meaningful.

The third phase (D) consists of degenerate configurations. These configurations are degenerate in the sense that they all correspond to the same value of $u_0 \rightarrow 1_-$ and are distinguished by different values of $w > 1.2$. Strictly speaking, a configuration of type $D$ is not a solution of the Oppenheimer-Volkoff equations, because no static solution     exists if $u_0 = 1$. In a solution with $u_0 = 1$, the bounding box would lie in a null (rather than a spacelike) hypersurface, thus violating the set-up of the system. The D-phase is a boundary of the thermodynamical state space that is obtained from solutions to the Oppenheimer-Volkoff equations through a limiting procedure. In practice, the introduction of a cut-off length-scale $l$, restricting the possible values of $u_0$ to be smaller than $1 - 2l/R$, allows for an identification of the global maxima of entropy with the solutions corresponding to $u_0 = 1 - 2l/R$, at least for values of temperature $T$ such that $l T  << 1$.

We believe that the degenerate solutions should be physically interpreted as a phase of complete gravitational collapse. The idealization of a box bounding self-gravitating radiation becomes unphysical, because for solutions at $u_0 \rightarrow 1$ the stresses on the box  tend to infinity. We do not obtain an actual black hole solution, because the  equation of state of radiation does not admit solutions with vanishing density at finite radius, hence, it cannot describe compact objects, unbounded by a box. We expect that systems with different equations of state have a richer and more physically relevant phase structure at the corresponding limit.

In Fig. 8, we plot the function  $\sigma_{eq}(w) = S_{eq}/R^{3/2}$, where  $S_{eq}(U,R)$ is the entropy of equilibrium states, defined by Eq. (\ref{seq}). For regular solutions, $\sigma_{eq}$ is an increasing function of $w$; for singular solutions $\sigma_{eq}$ remains an increasing function  of $w$ up to $w \simeq 0.51$ where it starts decreasing and  vanishes at $w = 1.2$. Degenerate solutions have vanishing entropy.

  Fig. 8 shows that for any given `radius' $R$ there is a maximal value of $S_{tot}$ that approximately equals  $0.823 R^{3/2}$. This contrasts the behavior of entropy in non-gravitating systems, where the entropy remains an increasing function of the internal energy at constant volume and has no maximum. On the other hand, the existence of a maximal value for the equilibrium entropy at constant $R$  is in accordance with the idea of an entropy bound for localized  gravitating systems \cite{bound}.

  There is also a global minimum to $S_{tot}$ for fixed $R$, approximately equal to $-1.65 R^{3/2}$. Since we chose the scale of entropy such that $S_{tot} = 0$ for the solutions with $M=0$, the positivity of entropy suggests a bound  on the gravitational entropy $S_{flat}(R)$ of flat space within a box of radius $R$, namely,  $S_{flat}(R) \geq 1.65R^{3/2}$. This bound is not particularly restrictive. For example, any expression for $S_{flat}(R)$ of the form $c R^{\gamma}$, for $\gamma > 3/2$ and $c$ of order unity satisfies this bound for all radii $R$ larger than the Planck scale.

\begin{figure}[tbp]
\includegraphics[height=6cm]{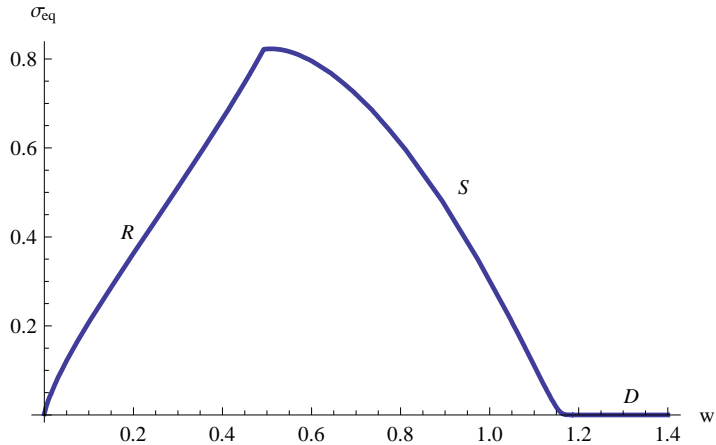} \caption{ \small  $\sigma_{eq}$ is plotted as a function of $w$. The first derivative of $\sigma_{eq}$ is discontinuous at the value of $w=u_Q$ corresponding to the transition from regular to singular solutions, but $\sigma_{eq}$ remains an increasing function until $w \simeq 0.51$. Then $\sigma_{eq}$ decays to zero with increasing $w$. Degenerate solutions all have zero entropy.}
\end{figure}

The decrease of entropy with increasing internal energy $U$ at constant radius $R$, characterizing the S-phase, would be quite unphysical in ordinary thermodynamics, because it would imply a negative temperature. However, in a self-gravitating system the equilibrium configuration is inhomogeneous and there is no guarantee that the variable $T$, corresponding to the temperature at the bounding box, has a simple relation to the entropy.

In the regular phase, there exists only a single boundary, and  the  relation between temperature  and entropy is formally analogous to the one in extensive thermodynamics, namely,
$T_{\infty}^{-1} = (\partial S_{eq}/\partial U)_R$ \cite{PWZ}. This simple relation is lost in the singular phase, because singular solutions
are   characterized by an internal boundary at the singularity. In fact, the singular solutions exhibit a feature very much unique to gravitational systems: the boundary of the system consists of two disconnected components, namely, the box at $r = R$ and the singularity at $r = 0$. The two components of the boundary have different temperature, and yet the system is in equilibrium with no energy flow. For this reason, the relation between the temperature(s) of the system and the derivatives of entropy is expected to be much more complex. We have not been able to obtain a physically intuitive form of this relation. The study of other gravitating system with an inner boundary is necessary, in order to better understand this issue.


The results obtained above for the non-equilibrium singular solutions rely on the identification of the internal energy $U$ with the Komar mass of the solutions.The Komar mass for radiation equals $2 \int_{\Sigma} \sqrt{-g} d^3x \rho$, where $\Sigma$ is a Cauchy surface, i.e., it includes only the contribution of radiation. In principle, it is possible that the singularity contributes an additional  term $U_{sing}$ defined from the values of the field variables at $r = 0$, analogous to the term $S_{sing}$ in total entropy.

 We have found no argument that allows us to uniquely determine   such a term. The reason is that the key thermodynamic properties are not significantly  affected by the choice of $U$. In particular, the proof in Sec. 3.3, that the regular solutions   maximize the total entropy for fixed $U$ and $R$, holds for any  choice of  $U \geq M$.

 The simplest choice for $U_{sing}$ is a term proportional to $R T_{\infty}S_0$, as in Eq. (\ref{uuu}). This means that the  internal energy is of the form
 \begin{eqnarray}
 U = M + \frac{\lambda}{2} R T_{\infty}S_0,
 \end{eqnarray}
 for some free parameter $\lambda$.
We have verified numerically that the curves of Fig. 7 remain unchanged for a large range of values of $\lambda$ (roughly, for $\lambda \in [0, 10]$), i.e., the entropy maxima at constant $U$ are insensitive to the choice of $\lambda$, even for the singular solutions. Hence, the only effect of a different definition of $U$ is a reparameterization of the variables $w$ in the plots of Fig. 7; in particular, the decrease of the equilibrium entropy in the S-phase remains unchanged.

\subsection{Phase transitions}

Next, we examine the transitions between the different phases R, S, and D of self-gravitating radiation.
In systems with long-range forces, the usual arguments about the equivalence of the microcanonical and the canonical distribution (at the thermodynamic limit) fail to apply \cite{ineq}. Hence, the phase structure of the equilibrium states of an isolated system may well differ from that of a system in contact with a thermal reservoir at constant temperature $T$.

\smallskip

We first consider an isolated box containing self-gravitating radiation. The relevant thermodynamic potential is the equilibrium entropy $S_{eq}$. The phase structure is  described by Fig. 8. There,  we note two transition points, one at $w = u_Q$ corresponding to the transition from the R-phase to the S-phase, and another at $w = 1.2$ corresponding to the transition from the S- phase to the D-phase. Both phase transitions are continuous.

At $w = u_Q$, the first derivative of $S_{eq}$ is discontinuous: we find that $\left(\frac{\partial S_{eq}}{\partial U}\right)_+ - \left(\frac{\partial S_{eq}}{\partial U}\right)_- \simeq 3.2 R^{1/2}$, where $+$ refers to $w \rightarrow u_Q$ from above (S-phase) and $-$ to $w \rightarrow u_Q$ from below (R-phase).

The first derivative of the temperature $T$ at $w = u_Q$ has an infinite discontinuity \cite{PL}. We find that $\left(\frac{\partial T}{\partial U}\right)_- \simeq -0.57R^{-3/2} (u_Q - w)^{-1}$ as $w \rightarrow u_Q$ and $\left(\frac{\partial T}{\partial U}\right)_+ \simeq -0.28R^{-3/2}$. Thus, there is a finite discontinuity of the heat capacity $C = \frac{\partial U}{\partial T_{\infty}}$ at $w = u_Q$: $C_+ - C_- \simeq - 5.05 R^{3/2}$. Numerical evaluation has not shown any discontinuities in the first derivatives of the basic thermodynamic variables at the second critical point $w = 1.2$.

Since the singularity entropy $S_{sing}$ vanishes for the regular solutions and it is non-zero for the singular and degenerate solutions, it is natural to associate $|S_{sing}|$ with the order parameter for the system. We find that near the critical point $w = u_Q$, $|S_{sing}| \sim 1.73 R^{3/2} (w - u_Q)$. The behavior of the system near the critical point suggests a phase transition described by the critical exponents of mean-field theory (with $\sqrt{|S_{sing}|}$ as the order parameter), as is to be expected from a treatment that relies on the maximum-entropy principle.

\smallskip

 Next, we consider the box in contact with a thermal reservoir at temperature $T$. As   seen from Fig. 7, there exists a range of values of $v_0$, and hence a range of values of $T$,   where all three phases coexist. The equilibrium phase corresponds to the global minimum of the Helmholtz free energy $F = U - T_{\infty} S_{eq}$. As in ordinary equilibrium thermodynamics, this follows from the entropy maximization of the total system including the box with radiation and the thermal reservoir, and taking into account that an exchange of energy $\delta Q$ at the boundary corresponds to a change $\delta U = \sqrt{1 - u_0} \delta Q$ of the internal energy of the box. In Fig. 9, the scaled Helmholtz free energy $f = F/R^{3/2}$ is plotted as a function of the scaled local temperature $TR^{1/2}$.

\begin{figure}[tbp]
\includegraphics[height=7cm]{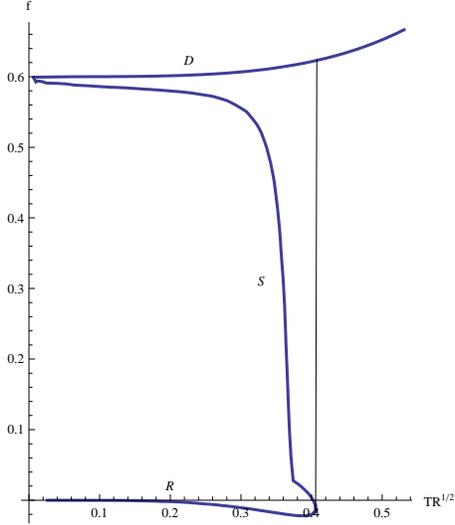} \caption{ \small  The scaled Helmholtz free-energy $f = F/R^{3/2}$ is plotted as a function of the scaled local temperature $TR^{1/2}$ for all equilibrium solutions.}
\end{figure}

The regular phase corresponds to a global minimum of the free energy for all values $TR^{1/2} \in [0, 0,406)$. For higher temperatures only the degenerate phase exists. The critical point $TR^{1/2} \simeq 0.406$ corresponds to the point $P$ of Fig. 1. The segment $PQ$ of Fig. 1 as well as the S-phase do not correspond to global minima in the free energy, and thus they do not define  stable equilibrium states. Hence, when the box is in contact with a thermal reservoir, there is a single critical point where the system undergoes a {\em first-order} phase transition, from the regular to the degenerate phase. The latent heat for this transition (as measured by an observer at infinity) is defined as $L := U_D - U_R$, where $U_D$ and $U_R$ are the internal energies at the critical point for the D-phase and the R-phase respectively. We find that   $L \simeq 0.237 R$.

 The equilibrium states determined in Fig. 9 are characterized by non-negative values of $\frac{\partial U}{\partial T}$ ($\frac{\partial v_0}{\partial w} \geq 0$). This is in marked contrast to the case of the isolated box considered earlier. This property reflects the positivity of the heat capacity characterizing systems described by the canonical distribution \cite{Padm}.

The phase structure of self-gravitating radiation is qualitatively similar to the phase structure of isothermal Newtonian spheres, regularized at small scales---see, Ref. \cite{Cha2} for a review of such transitions using the microcanonical and canonical ensembles. We note that the approach we employ in this paper provides only  a rough approximation to the physics of the phase transition. The reason is that the maximum-entropy principle cannot account for the effect of thermodynamic fluctuations that are essential for the precise description of critical phenomena \cite{Callen}.  A more reliable approach for the study of phase transitions should involve   statistical thermodynamics. The entropy functional $S_{tot}$ allows the definition of
 equilibrium states as averages with respect to the probability distribution $\exp [ S_{tot}(U, T, R)]$. Thus, the contribution of the fluctuations near the critical points would be incorporated  into the thermodynamic description.

 \subsection{Relation to Bekenstein-Hawking entropy}

 Black hole entropy has provided the most important indication of a fundamental relation between gravity and thermodynamics. Therefore, it is natural to inquire whether the entropy of singularities, proposed here, is related to the Bekenstein-Hawking entropy.

 The arguments developed in this paper  involve only the basic principles of general relativity and thermodynamics. At no point has it been necessary to introduce  quantum theory, and, consequently $\hbar$ does not appear in any of the expressions. (It only appears implicitly through the  Stefan-Boltzmann constant.) Hence,  the Bekenstein-Hawking entropy cannot appear in the present context, because it depends explicitly on $\hbar$,.
  The singularities of self-gravitating radiation have less entropy than locally Minkowski
  an spacetime, while black hole horizons have much larger entropy. Presumably, they originate from  different sectors in the  space of gravitational microstates.

 In Sec. 2.3.1, we noted that the system admits a class of trivial solutions for $u_0 \in (0, 1)$ and $v_0 = 0$ that correspond to a Schwarzschild horizon of radius $r_S = 2M$  inside the box and no radiation. These solutions form another, fourth, phase of the system.  Within the classical theory, their entropy is zero; hence, by the maximum-entropy principle they do not define equilibrium solutions. However, one might inquire whether an assignment of Bekenstein-Hawking entropy $S_{BH} = \pi r_S^2/\hbar$ to these solutions can be consistently incorporated into the present framework. The answer is negative for the following reasons.

  A black hole horizon, endowed with thermodynamic properties,  is expected to emit Hawking radiation. The presence of a bounding box implies that this radiation does not escape to infinity. Thus, the equilibrium configuration should correspond to the horizon coexisting with its Hawking radiation.  The local temperature $T$ at the  bounding box would equal  the Hawking temperature blueshifted by a factor of $\sqrt{1 - u_0}$, as   in Ref. \cite{york}. But, as we showed in Sec. 2.3, there is no solution of the Oppenheimer-Volkoff equations with both non-zero temperature at the boundary and an horizon.  We would have to modify the constitutive equations of the system in order to introduce the Bekenstein-Hawking entropy into the present set-up. Such modifications could arise from the consideration of quantum corrections to Einstein's equations, or from the inclusion of the effect of acceleration radiation, or from modifications to the equation of state for radiation near the horizon \cite{mod}.  However, any   modification would   affect the entropy of the system (both of radiation and of the singularities), possibly introducing additional terms with different scaling properties. As a matter of fact, such considerations might provide an important generalization of the results presented here, namely, the determination of constitutive equations that incorporate the Bekenstein-Hawking entropy into the classical thermodynamic description.


\section{Conclusions}

We have argued that the requirement of thermodynamical
 consistency, when applied to self-gravitating systems, can provide novel information about the thermodynamic properties of the gravitational field. We applied this idea to a concrete system, namely, static, spherically symmetric solutions to Einstein's equations describing self-gravitating radiation in a box. We showed that the only way to describe this system in a thermodynamically consistent way requires the assignment a specific expression $S_{sing}$ to the entropy of the spacetime singularities. While this result is restricted to a   specific class of self-gravitating system, it does demonstrate a point of fundamental importance: that horizons are not the only entities of gravity theory to which entropy can be assigned. This implies the intrinsic connection between gravity and thermodynamics is not restricted in the  black hole context.

 Of equal importance is the fact that the expression $S_{sing}$ for the singularity entropy is identified through stringent constraints of physical and mathematical consistency. The existence of a functional satisfying such constraints is a highly non-trivial result. It suggests that the
  thermodynamic description of the gravitational field is fundamentally embedded in Einstein's equations, and the rationale developed here can be applied to other physical systems. In particular, we  expect that the justification of the regularity condition through  maximum-entropy principle holds in general spherically symmetric solutions to Einstein's equations, as long as the equations of state are thermodynamically consistent. The study of a large class of such systems will enable the solidification of this hypothesis.  The eventual aim is to identify gravitational entropy within the broadest framework where the principles of equilibrium thermodynamics are applicable, namely, for general static solutions to Einstein's equations.

  The main limitation of the present results is the fact that they are based on numerical solution of the Oppenheimer-Volkoff equations, rather than a full analytic derivation. The application of techniques from the theory of differential equations is an important requirement of further research, because it can provide a firmer analytic control on the properties of singular solutions. The definition of the singularity entropy requires very specific behavior of the fields near the singularity. It is essential to understand the mathematical origins of this behavior, in order to proceed towards directions of increasing generality.

It is important to emphasize that the methodology developed here employs only assumptions from classical theories, namely, general relativity and thermodynamics. Thus, any results obtained from this method can, in principle, be  employed in order to constrain candidate quantum theories of gravity.  The study of phase transitions is perhaps the most important in this direction.  As explained in Sec. 3.5, the maximum-entropy principle is not adequate for the description of the systems near the critical point. However, the definition of a consistent entropy functional provides the basis for a treatment of the critical behavior using statistical thermodynamics. If the gravitational phase transitions exhibit universality, then the identification of a universality class could provide novel information about the symmetries of the underlying microscopic theory.

This paper is partly motivated by a change of perspective towards singularities. We do not view them as pathologies of classical general relativity, to be resolved by a quantum theory of gravity,  but as indications of a different thermodynamic phase for the gravitational field, which appears when the external constraints are not compatible with locally Minkowskian geometry. This perspective might have implications in the theory of stellar stability, where physical solutions are standardly selected by the regularity assumption.
 A thermodynamic treatment of systems with  realistic equations of state, could uncover a new phase
of singular equilibrium solutions, similar to the S-phase appearing in self-gravitating radiation. If such a phase turns out to be thermodynamically stable when the system is in contact with  heat and/or pressure reservoirs, it would suggest the existence of a new type of compact astrophysical objects, that interpolates between ordinary stars (regular solutions to Einstein's equation) and black holes.


\begin{thebibliography}{}

\bibitem{Jac} T. Jacobson, Phys. Rev. Lett. 75, 1260 (1995).

\bibitem{Pad} T. Padmanabhan, Gen. Rel. Grav. 34, 2029 (2002); Phys. Rept. 406, 49 (2005); A.Mukhopadhyay and T. Padmanabhan, Phys.Rev.D74,124023 (2006); T. Padmanabhan, Rep. Prog. Phys. 73, 046901 (2010).

\bibitem{Ver} E. P. Verlinde, arXiv:1001.0785.

\bibitem{Hu} B. L. Hu, arXiv:1010.5837.

\bibitem{Pen} R. Penrose, in  {\em Einstein Centenary Volume}, S. W.  Hawking and G. Ellis (eds.), (Cambridge Uni-
versity Press, 1979); R. Penrose, in {\em The Future of Theoretical Physics and Cosmology} (Cambridge University Press, 2002).

\bibitem{PWZ} R. D. Sorkin, R. D. Wald and Z. Z. Jiu, Gen. Rel. Grav. 13, 1127 (1981).

\bibitem{Cha1} P.H. Chavanis, A\&A, 483, 673 (2008).

\bibitem{Padm} T. Padmanabhan, Phys. Rep. 188, 285 (1990).

\bibitem{Katz} J. Katz, Found. Phys. 33, 223 (2003).


\bibitem{Cha2} P.H. Chavanis, Int. J. Mod. Phys. B, 20, 3113 (2006).

\bibitem{CDR} A. Campo,  T. Dauxois and S. Ruffo, Phys. Rep. 480, 57 (2009).


\bibitem{Opp} J. Oppenheim, Phys. Rev. D65, 024020 (2001); Phys. Rev. E68, 016108 (2003).

\bibitem{Pes} A. Pesci, Class. Q. Grav. 24, 2283 (2007).


\bibitem{LL} L. Landau and E. M. Lifshitz, {\em Statistical Physics}, (Pergamon Press, New York), pg. 21.

\bibitem{Thi}    W. Thirring, Z. Physik 235, 339 (1970).


\bibitem{ineq} P. Hertel and W. Thirring, Ann. of Phys. 63, 520 (1971); D.H.E. Gross and E. Votyakov, Eur. Phys. J. B15, 115 (2000).


    \bibitem{gth} V.A. Antonov, Vest. Leningrad Gros. Univ. 7, 135 (1962);  D. Lynden-Bell and R. Wood, Mon. Not. RAS 138, 495 (1968); S. Tremaine, M. Henon and D. Lynden-Bell, M. N. R. A. S. 219, 289 (1986); T. Padmanabhan, Astr. J. Suppl.  71, 651 (1989).


\bibitem{Gil} R. Giles, {\em Mathematical Foundations of Thermodynamics}, (Pergamon, Oxford, 1964).

\bibitem{LY} E. H. Lieb and J. Yngvason, Phys. Rep. 310, 1 (1999).




    \bibitem{Callen} For a presentation of thermodynamics emphasizing the role of the maximum-entropy principle, see, H. B. Callen, {\em Thermodynamics and an Introduction to Thermostatistics} (John Wiley, New York, 1985).

\bibitem{oth} One needs not invoke the full force of the 0-th law of the thermodynamics for this statement, which fails in systems with long-range forces---see, Ref. \cite{Thi} and A. Ramirez-Hernandez, H. Larralde and F. Leyvraz, Phys. Rev. Lett. 100, 120601 (2008). It suffices that the points of contact with the reservoir acquires the reservoir's temperature. The local temperature of the self-gravitating gas varies with the distance from the bounding box.

\bibitem{PL} D. Pavon and P. T. Landsberg, Gen. Rel. Grav. 20, 457 (1988).


\bibitem{ST} J. Smoller and B. Temple, Arch. Rational Mech. Anal. 142, 177 (1998).

\bibitem{thooft} G. 't Hooft, Nucl. Phys. B256, 727 (1985); Int. J. Mod. Phys. A11, 4623 (1996).

\bibitem{ZP} W. H. Zurek and D. N. Page, Phys. Rev. D29, 628 (1984).

\bibitem{JH} S. Deser, R. Jackiw and G. 't Hooft, Ann. Phys. 152, 220 (1984).

\bibitem{cch} R. Penrose, in {\em Black Holes and Relativistic Stars}, R. M. Wald (ed.) (University of Chicago Press, 1998); R. M. Wald, gr-qc/9710068.

\bibitem{Sred} M. Srednicki, Phys. Rev. Lett. 71, 666 (1993)

\bibitem{GM} R. S. de Groot and P. Mazur, {\em Non-Equilibrium
Thermodynamics}, (Dover, 1984).

\bibitem{bound} J. D. Bekenstein, Phys. Rev. D23, 287 (1981); Phys. Rev. D49, 1912 (1994); Phys. Rev. D60, 124010 (1999); R. Bousso, JHEP 9906, 028 (1999); Rev. Mod. Phys. 74, 825 (2002); R. Brustein and G. Veneziano, 	Phys. Rev. Lett. 84, 5695 (2000);
    R. Bousso, E. E. Flannagan and D. Marolf, Phys.Rev. D68, 064001 (2003).

\bibitem{york} J. W. York, Phys. Rev. D 33, 2092 (1986); B. F. Whiting and J. W. York, Phys. Rev. Lett. 61, 1336 (1988).

\bibitem{mod} W. G. Unruh and R. M. Wald, Phys. Rev. D25, 942 (1982); L. Li and L. Liu, Phys. Rev. D46, 3296 (1992); W. G. Anderson,  Phys. Rev. D50, 4786 (1994); D. X. Wang, Phys. Rev. D50,  7385 (1994); D. X. Wang, Gen. Rel. Grav. 27, 1251 (1995).


\end{thebibliography}
\end{document}